\documentclass[fleqn,twoside]{article}
\usepackage{amsmath,amssymb,psfrag}
\usepackage[center]{subfigure}
\usepackage{espcrc2}

\usepackage{graphicx}
\usepackage{color}
\definecolor{refkey}{rgb}{0.5,0.5,0.5}
\definecolor{labelkey}{rgb}{0.8,0.8,0.8}

\definecolor{gray}{rgb}{0.6,0.6,0.6}
\usepackage{feynmp}

\newcommand{\psibar}{{\overline{\psi}}}
\title{
Further Improvements to Staggered Quarks}
\author{Eduardo Follana\address[Glasgow]{Department of Physics \& Astronomy, The University of Glasgow, Glasgow G12 8QQ, UK}, Quentin Mason\address[Cornell]{Laboratory of Elementary-Particle Physics, Cornell University, Ithaca, NY 14853, USA}\thanks{Presented at Lattice 2003, current address: DAMTP, Cambridge University, UK}, Christine Davies\addressmark[Glasgow], Kent Hornbostel\address[SMU]{Department of Physics, Southern Methodist University, Dallas, TX 75275, USA}, Peter Lepage\addressmark[Cornell] and Howard Trottier\address{Physics Department, Simon Fraser University, Burnaby, B.C.  V5A 1S6, Canada}, HPQCD collaboration.}

\begin{document}
\begin{abstract}
The improved staggered quark action has enabled breakthrough calculations of QCD with three light dynamical quarks.  The largest remaining $a^2$ scaling violations come from ``taste-changing'' interactions, which are significantly reduced over na\"\i ve quarks because of carefully chosen smearing of the gauge links.  Here we examine further improvements to the quark action and show that additional smearing of the links can reduce the taste-changing errors by another factor of two, whilst retaining a fully $a^2$-improved action.  The taste-changing errors can be understood within a perturbative calculation of the four-quark operators that appear at one-loop.  We also discuss unquenching techniques with such a highly improved action.
\vspace{-1em}
\end{abstract}
\maketitle
Accurate lattice simulations of the Standard Model require numerical simulations in lattice QCD which include three light flavours of dynamical quarks.  The improved staggered-quark (Asqtad) formalism is the only one capable of delivering large numbers of configurations with three small quark masses anytime in the near future~\cite{Jansen}.  Recent calculations~\cite{Davies:2003ik} have shown that such a formalism is already highly successful, delivering numbers for ``gold-plated'' quantities that agree with experiment at the few percent level.  This analysis, while impressive, highlighted that increased efficiency in the simulations would be possible if the action could be further improved.  Here we systematically explore this possibility.

Staggered quarks have 16 non-degenerate pions made from the four ``tastes'' --- the remnant of the doublers --- whose masses do \emph{not} vanish for zero quark mass.  The splitting comes from mixing between the tastes and vanishes like $a^2$.  Na\"\i ve staggered quarks suffer from poorly convergent perturbative expressions and large pion splittings which are suppressed by a factor of 2--3 for typical simulations by the use of fat links~\cite{Blum:1996uf}, consistent with the reduction of a pure $a^2$ effect to $\alpha_s a^2$.  Staggered quarks which are improved to $\mathcal{O}(\alpha_s a^2,a^4)$-errors~\cite{Lepage:1998vj}, significantly reduce splittings in the pion spectrum~\cite{Toussaint:2001zc}, and have small renormalisations~\cite{Hein:2001kw}.  A related scheme with smaller pion splittings is the HYP action~\cite{Anna}.

We can further reduce these undesirable splittings by adding explicit 4-quark counter-terms to the lattice action~\cite{boston}.  These are computed in 1-loop perturbation theory (since taste-changing is perturbative).  As we show, a simple alternative is to suppress the 4-quark counter-terms by modifying the quark-gluon vertices.  Using insight drawn from perturbation theory, we have designed a new quark action that has comparable taste-changing interactions to HYP, but is much simpler to implement and unquench and retains the essential $a^2$-improvement necessary for accurate simulations.

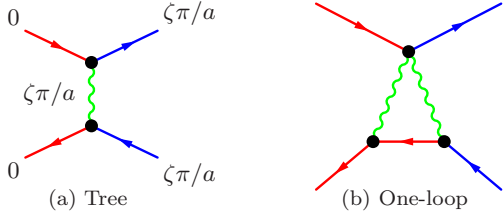
\begin{figure}[t]%
\begin{center}%
\renewcommand\subfigcapskip{0pt}
\renewcommand\subfigbottomskip{0pt}
\renewcommand\subfigcapmargin{0pt}
\subfigure[Tree\label{f:tree}]{\;%
    \begin{fmfgraph*}(50,60)%
      \fmfset{arrow_len}{2mm}%
      \fmfset{wiggly_len}{2.5mm}%
      \fmftopn{l}{2}%
      \fmfbottomn{r}{2}%
      \fmfv{lab=\small $0$,lab.dist=3pt}{l1}%
      \fmfv{lab=\small $\zeta\pi/a$,lab.dist=3pt}{l2}%
      \fmfv{lab=\small $0$,lab.dist=3pt}{r1}%
      \fmfv{lab=\small $\zeta\pi/a$,lab.dist=3pt}{r2}%
      \fmf{fermion,f=red}{l1,v1}
      \fmf{fermion,f=blue}{v1,l2}%
      \fmf{fermion,f=blue}{r2,v2}%
      \fmf{fermion,f=red}{v2,r1}%
      \fmf{boson,f=green,label=\small $\zeta\pi/a$}{v1,v2}%
      \fmfdot{v1,v2}%
    \end{fmfgraph*}}%
\qquad\qquad\quad
\subfigure[One-loop\label{f:oneloop}]{%
{\raisebox{-15pt}[50pt][0pt]{\quad\begin{fmfgraph}(70,88)
  \fmfset{arrow_len}{2mm}%
  \fmfset{wiggly_len}{2.5mm}%
  \fmftopn{l}{2}%
  \fmfbottomn{r}{2}%
  \fmf{fermion,f=red,tens=1.5}{l1,v1}%
  \fmf{fermion,tens=1.5,f=blue}{v1,l2}%
  \fmf{fermion,tens=1.5,f=blue}{r2,v2}%
  \fmf{fermion,f=red,tens=1.5}{v3,r1}%
  \fmf{fermion,f=red,tens=0.8}{v2,v3}%
  \fmf{boson,f=green,tens=0.8}{v2,v1,v3}%
  \fmfdot{v1,v2,v3}%
\end{fmfgraph}
}}}
\vspace{-5mm}
\caption{\label{f:tastechange}Sample taste-changing diagrams for massless na\"\i ve quarks}%
\end{center}%
\end{figure}

Taste-changing is most easily understood in terms of na\"\i ve quarks, which are equivalent to 4 identical staggered quarks.  Na\"\i ve quarks have an exact symmetry:
\begin{equation}
  \text{quark}(p\sim 0)\quad\equiv\quad \text{quark}(p\sim\zeta\pi/a)
\end{equation}
where $\zeta=(1,0,0,0)$, $(1,1,0,0)$, \ldots\; giving 16 identical copies or ``tastes'' of every quark, one in each corner of the Brillouin zone.  These different tastes can mix by exchanging a hard gluon as shown in figure~\ref{f:tree}.  This gluon is highly virtual with momentum $\mathcal{O}(\pi/a)$ and thus the quark-quark interaction is effectively a purely perturbative contact interaction at typical lattice spacings.  
\begin{fmffile}{japan_taste_feyn}%
The tree-level interaction of figure~\ref{f:tree} was understood and completely removed with the improved staggered action by using smearing to suppress high-momentum gluon emission from quarks~\cite{Blum:1996uf,Lepage:1998vj}.  At one-loop order, there are 5 non-zero diagrams that contribute to the taste changes; one is shown in figure~\ref{f:oneloop}.  In the massless limit there are 21 independent counter-terms coming from 1-loop diagrams:
\begin{align}
 \Delta\mathcal{L} = \sum_{\zeta}\frac12 c_\zeta J_\zeta^2.
\end{align}
It is not obvious which of these terms is most important, but a crude estimate of the taste-changing is given by $\sum|c_\zeta|$.  We want to minimise this quantity when designing new, further improved, quark actions.

One-loop taste-changing requires large momentum transfers through the gluons, consequently we can suppress these contributions by softening the quark-gluon vertices at high momentum.  This softening can be achieved by carefully smearing \emph{all} of the link operators in the quark action.  

Extensive smearing leads to a large number of terms, many of which involve very long products of link operators.  While 1-gluon vertices are well behaved by design, 2-gluon vertices ($\psibar A^2\psi$) tend to become huge and result in large taste-changing contributions.  This problem can be minimised by re-unitarising the smeared-link operators.
\begin{figure}[t]
\begin{center}
\psfrag{U}[][b]{\small Fat7}
\psfrag{V}[][b]{\small Fat7$\otimes$Fat7}
\psfrag{R}[br]{}
{\qquad\quad\includegraphics[scale=0.4]{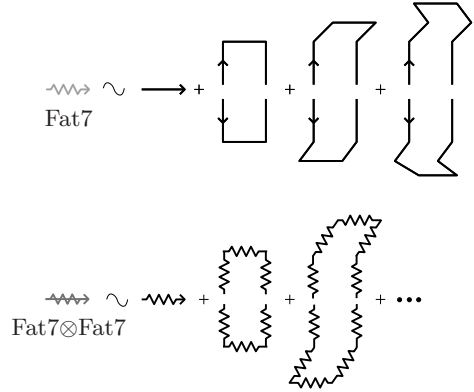}}
\caption{A schematic of Fat7 smearing, and the principle behind smeared-smearings: the original smeared links are used in the second smearing.  The $\sim$ is because appropriate coefficients and sums over all directions of the path elements are not shown.\label{f:fat7smearing}}
\end{center}
\end{figure}

\begin{figure*}[!t]
\begin{center}
\psfrag{alongtitleatthetop}[cc][c]{Pion Splitting in Simulation and Perturbation Theory}
\psfrag{k1aaaaaaaaaaaa}[Bl][Bl][0.85]{\small Simulation}
\psfrag{k2aaaaaaaaaaaa}[Bl][Bl][0.85]{\small Perturbative}
\psfrag{deltam2}[][r][1][-90]{$\Delta m_\pi^2a^2$}
\psfrag{sumc2}[Bl][t][1][-90]{$\displaystyle\sum_\zeta |c_\zeta|$}
\psfrag{A1}[lt][c][1][-22]{Asqtad}
\psfrag{A2}[lt][c][1][-22]{Fat7}
\psfrag{A3}[lt][c][1][-22]{HYP}
\psfrag{A4}[lt][c][1][-22]{Fat7$_{R}\otimes$Fat7$_{R}$}
\psfrag{A5}[lt][c][1][-22]{{Fat7$_{R}\otimes$Asqtad}}
{\includegraphics[scale=0.38,angle=270]{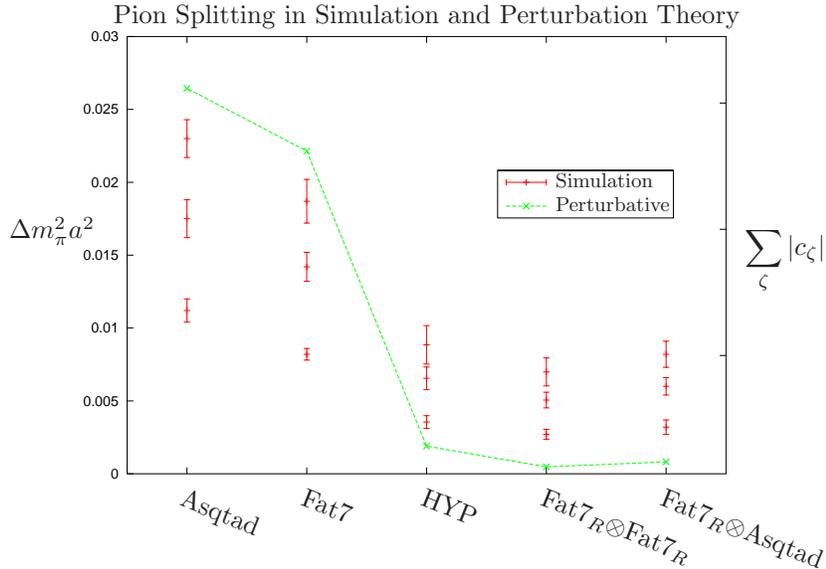}}
\vspace{0.5cm}
\caption{Quenched, $\beta=5.93$, $m=0.03$ simulations with Wilson glue of the splittings in the squared masses of 1-, 2- and 3-link pions above the Goldstone.  Errors are statistical, and come from uneven numbers of configurations.  The perturbative estimate of the taste-changing shown on a different scale qualitatively tracks the simulation data.  The proposed action Fat7$_{R}\otimes$Asqtad, shown at the right, is about two times better than Asqtad, shown on the left.\label{f:finalpic}}
\end{center}
\end{figure*}

Tree-level taste-changing was removed by ``Fat7'' smearing (figure~\ref{f:fat7smearing})~\cite{Lepage:1998vj}, but this leaves the side links unsmeared.  A better choice is to apply Fat7 twice: Fat7$\otimes$Fat7 (figure~2b).  To avoid the combinatorial growth of $\psibar A^2\psi$ vertices, the links are re-unitarised after each smearing: Fat7$_R\otimes$Fat7$_R$.  The $a^2$-improved cousin, Fat7$_R\otimes$Asqtad, uses the same re-unitarised Fat7-smeared links in the original Asqtad action, but with only one re-unitarisation step.  We also investigated SU(3) back-projection instead of re-unitarisation, however, it is much harder to unquench for no discernible benefit.  We have developed two methods to compute the unquenched molecular dynamics force term~\cite{R}.  Re-unitarisation, $V_R\equiv (1/\sqrt{VV^\dagger}) V$, can be differentiated via the chain rule~\cite{Kamleh}.  This can be done explicitly or approximated stochastically, the two methods give consistent results but the stochastic estimation, while introducing additional noise, is faster.  The Fat7$_R\otimes$Fat7$_R$ action, like the HYP action, has tree-level $a^2$ errors.   These are absent in the Fat7$_R\otimes$Asqtad, and this action has $\sim2$ times smaller taste-changing than the traditional Asqtad action.  This suggests that Fat7$_R\otimes$Asqtad is the most highly improved action to date.  We compare the trend in $\sum|c_\zeta|$, computed perturbatively, with quenched simulation results for the splitting in squared pion mass in figure~\ref{f:finalpic}.  The perturbative results track the simulation results, confirming that taste-changing is essentially perturbative in character.

These studies confirm that Fat7$_R\otimes$Asqtad is the most accurate light-quark action available, with residual discretisation errors of $\mathcal{O}(a^2\alpha_s)$, and taste-changing that is substantially smaller than for Asqtad and comparable to HYP.  Fat7$_R\otimes$Asqtad is far simpler to code than the HYP action, more accurate and easier to unquench. Finally, these studies demonstrate that taste-changing is essentially a perturbative phenomenon.  Significant further improvements, through more complex smearings, are probably difficult or impossible.

\subsection*{Acknowledgement}
We are grateful for assistance from Anna Hasenfratz in implementing the HYP action.
\end{fmffile}

\end{document}